\documentclass[useAMS,usenatbib]{mn2e}
\usepackage{amsmath}
\usepackage{threeparttable}
\usepackage{graphicx}
\usepackage{color}

\newcommand\mnras{MNRAS}%
\newcommand\aj{AJ}%
\newcommand\apj{ApJ}%
\newcommand\apjs{ApJS}%
\newcommand\aaps{A\&AS}%
\newcommand\pasp{PASP}%
\title[Matching Radio Catalogs with Realistic Geometry:Application to SWIRE and ATLAS]{Matching Radio Catalogs with Realistic Geometry: \\ Application to SWIRE and ATLAS}
\author[Fan et al.]{
Dongwei Fan$^{1}$\thanks{E-mail:dfan2@pha.jhu.edu},
Tam\'as Budav\'ari$^{2,3}$\thanks{E-mail:budavari@jhu.edu},
Ray P. Norris$^{4}$,
Andrew M. Hopkins$^{5}$\\
$^{1}$Key Laboratory of Optical Astronomy, National Astronomical Observatories, \\Chinese Academy of Sciences, 20A Datun Road, Beijing 100012, China\\
$^{2}$Department of Physics and Astronomy, The Johns Hopkins University, 3400 North Charles Street, Baltimore, MD 21218, USA\\
$^{3}$Department of Applied Mathematics and Statistics, The Johns Hopkins University, 3400 North Charles Street, Baltimore, MD 21218, USA\\
$^{4}$CSIRO Astronomy and Space Science, Epping, NSW 1710, Australia\\
$^{5}$Australian Astronomical Observatory, PO Box 915, North Ryde, NSW 1670, Australia
}
\begin{document}


\pagerange{\pageref{firstpage}--\pageref{lastpage}} \pubyear{2002}

\maketitle

\label{firstpage}

\begin{abstract}
Crossmatching catalogs at different wavelengths is a difficult problem in astronomy, especially when the objects are not point-like. At radio wavelengths an object can have several components corresponding, for example, to a core and lobes. {Considering not all radio detections correspond to visible or infrared sources, matching these catalogs can be challenging.} Traditionally this is done by eye for better quality, which does not scale to the large data volumes expected from the next-generation of radio telescopes. We present a novel automated procedure, using Bayesian hypothesis testing, to achieve reliable associations by explicit modelling of a particular class of radio-source morphology. {The new algorithm not only assesses the likelihood of an association between data at two different wavelengths, but also tries to assess whether different radio sources are physically associated, are double-lobed radio galaxies, or just distinct nearby objects.} Application to the SWIRE and ATLAS CDF-S catalogs shows that this method performs well without human intervention.
\end{abstract}


\section{Introduction}
To assemble time-series and multicolor data sets in astronomy, we routinely combine observations from different telescopes and instruments. Hundreds of catalogs are available today starting with the largest surveys \citep[e.g., SDSS SkyServer;][]{2000AJ....120.1579Y,2002cs........2013S}, to the many legacy observations in Vizier~\citep{viziercatalog}, which are served by Virtual Observatory (VO) services as well as Simbad~\citep{2000A&AS..143....9W} or the NASA Extragalactic Database \citep[NED;][]{2001SPIE.4477...20M}{, NASA/IPAC Infrared Science Archive~(IRSA)~\footnote{IRSA~{http://irsa.ipac.caltech.edu/}}}. These collections are accessible online and there are also specialised services that offer cross-identification.
Such services include the SkyQuery~\citep[]{skyquery, skyquery2}, CDS XMatch~\footnote{CDS xMatch service~{http://cdsxmatch.u-strasbg.fr/xmatch}} or the VAO Cross-Comparison Tool~\footnote{VAO Catalog Cross-Comparison Tool~{http://vao-web.ipac.caltech.edu/applications/VAOSCC/}}. These services use fast indexing methods, e.g., the Hierarchical Triangular Mesh \citep[HTM;][]{2001misk.conf..631K,budavari2010}, HEALPix~\citep{healpix} or the Zones Algorithm~\citep[]{2004cs........8031G,2007cs........1171G} that quickly eliminate sources too far away from each other.

These tools, however, are essentially designed for crossmatching between point sources. They are good for merging optical catalogs of stars and small extended sources but mostly useless for associating the largest galaxies or radio sources.
%
{For example, a single radio galaxy can have several distinct components corresponding to its core and lobes.}
Traditional methods can quickly link the radio core to the optical center, but the jet or lobes would be ignored since they are far away from the center. Currently such components are identified and associated by eye \citep[e.g.,][]{2006AJ....132.2409N}.

Matching by eye is adequate for existing deep radio surveys, with only a few thousand sources, but will be untenable for some surveys currently under way with tens of thousands of sources, and will be totally impractical for surveys such as EMU~\citep{2011PASA...28..215N} which is expected to detect 70 million sources. Existing attempts to build automated algorithms to perform the cross-matching of complex radio sources~\citep[e.g.,][]{2006ApJS..165...95P,2015MNRAS.446.2985V}, have used simple algorithms that will only identify the simplest cases.

Here we present an automatic method to estimate the matching probability for radio sources, which extends the Bayesian approach of \citet{2008ApJ...679..301B} to include an explicit geometric model. For the current analysis, this model is aimed at simulating a typical radio galaxy structure, composed of a core and two approximately symmetrically located lobes. In principle, though, any model can be used, and future work will generalise this to account for more complex radio source structures. Our approach is described in section~\ref{sec:model}. Its application and results are discussed in section~\ref{sec:app}. Section~\ref{sec:summary} presents our conclusions.

\section{Matching with Radio Morphology}\label{sec:model}
Probabilistic cross-identification has several advantages over traditional methods. Bayesian hypothesis testing uses individual estimates of the uncertainties in the directions of sources and provides an objective measure of the quality of a given association from the available data. It also just requires a single threshold value to define a matched catalog even if there are many input catalogs to consider. Its biggest advantage, however, is its flexibility to model the problem at hand. \citet{2010ApJ...719...59K} successfully applied the method to associate moving stars with unknown proper motion and \citet{2011ApJ...736..155B} extended the method to time-domain observations to find matching events, e.g., supernovae. Here we use the same concept to associate radio sources with an unknown number of components and geometry.

To decide whether a given association is a good match, the likelihoods of two hypotheses have to be calculated. In practice these are done as integrals over the parameter space of the model. For point sources this is a simple calculation with an analytic result \citep{2008ApJ...679..301B}, but in general it would have to be computed numerically.

Throughout this paper we use the flat-sky approximation local to each association. The point source's nearby coordinates are projected to its tangent plane. This operation converts the spherical geometrical problem to a classical plane geometrical issue. This approximation is very accurate for the small distances we consider and allows for fast calculations.

\subsection{Choosing a prior}
%
{
For this analysis we assume that a radio source complex is well modelled either by (in the case of a star forming galaxy or radio-quiet AGN) a single isolated object , or  (in the case of a radio-loud AGN) by a radio nucleus,
coincident with the host galaxy, together with a pair of lobes , resulting in a linear or near-linear arrangement of 2 or 3 components.
In some AGNs, either the nucleus or one of the lobes may be missing or below our detection limit.
In the absence of a detailed quantified distribution of radio source morphologies this is the simplest prior to assume, and appears to be valid for the vast majority of known radio sources.

A more comprehensive prior could be constructed from the distribution of radio sources found in existing survey data, such as FIRST\citep[]{1994ASPC...61..165B}, NVSS\citep[]{1998AJ....115.1693C}, SUMSS\citep[]{1999AJ....117.1578B,2003MNRAS.342.1117M}, ATLAS\citep[]{2006AJ....132.2409N}. This distribution would vary as a function of sensitivity and angular resolution, and so the construction of such a prior would be justified
only if the simpler approach taken here proved inadequate. However, we recognise that such a comprehensive prior may be required in subsequent implementations of this technique.}

\subsection{Straight-Line Model}
\begin{figure}
\begin{center}
\includegraphics[scale=0.35]{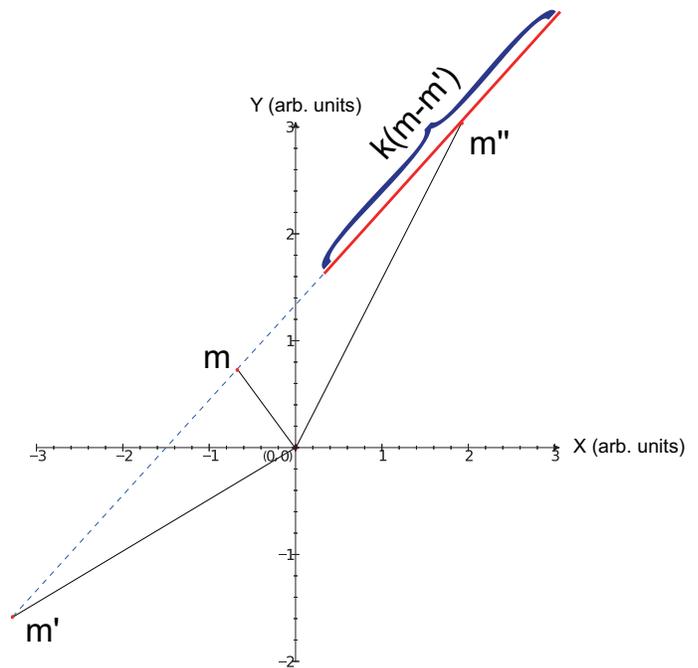}
\end{center}
\caption{A simple model of the radio morphology has a core and two lobes. The model coordinates are compared with the detections using the usual likelihood function based on a Gaussian astrometric uncertainty.\label{fig:model}
}
\end{figure}

As an example, we consider a simple triple radio source model, consisting of a radio core and two ejected lobes, equidistant from the core, with all three components lying along a straight line. The measured positions of the core and two lobes are denoted by the vectors $\boldsymbol{m}$, $\boldsymbol{m}^\prime$ and $\boldsymbol{m}^{\prime\prime}$ respectively, as shown in Fig.~\ref{fig:model}.
%
These are two-dimensional vectors on the tangent plane and, in this simplest case, obey the following relation
\begin{equation}\label{eqn:m22mm1}
\boldsymbol{m} = \frac{\boldsymbol{m}'+\boldsymbol{m}''}{2}
\end{equation}
which yields \mbox{$\boldsymbol{m}''\!=\!2\boldsymbol{m}\!-\!\boldsymbol{m}'$} as illustrated in Figure~\ref{fig:model}.

We must now compare this model with real data, consisting of the measured position of a putative optical host galaxy denoted by $\boldsymbol{c}$, and the measured positions of an assumed radio core and two lobes, denoted by $\boldsymbol{r}_0$, $\boldsymbol{r}_1$ and $\boldsymbol{r}_2$ respectively.
We label the measurements such that $\boldsymbol{c}$ and $\boldsymbol{r}_0$ correspond to the core at $\boldsymbol{m}$, while $\boldsymbol{r}_1$ and $\boldsymbol{r}_2$ are the measurements of the lobes at $\boldsymbol{m}'$ and $\boldsymbol{m}''$.
If the detections come from the same object with the core and lobes identified correctly, the likelihood of this hypothesis is calculated as the integral over $\boldsymbol{m}$ and $\boldsymbol{m}'$. The prior on $\boldsymbol{m}$ is just \mbox{$p(\boldsymbol{m})\!=\!1\big/4\pi$} as the object can be anywhere on the entire sky but the conditional probability density of $m'$ is localized around $m$, e.g., \mbox{$p(\boldsymbol{m}'|\boldsymbol{m})=f\big(|\boldsymbol{m}'\!-\!\boldsymbol{m}|\big)$}. We will study the effect of this choice later.
In this notation, the likelihood of the CORE-LOBE-LOBE configuration is
\begin{eqnarray}
\label{eqn:linelikelihood}
{\cal{}L}_{cll} &=& \left[\int\!d\boldsymbol{m}\ p(\boldsymbol{m})\,L_{c}(\boldsymbol{m})\,L_{r_0}(\boldsymbol{m})\right. \nonumber\\
			     && \left.\int\!d\boldsymbol{m}'\ p(\boldsymbol{m}'|\boldsymbol{m})\, L_{r_1}(\boldsymbol{m}')\,L_{r_2}(2\boldsymbol{m}\!-\!\boldsymbol{m}')\right]
\end{eqnarray}
where the likelihood functions have the usual form, say a Gaussian \mbox{$L_{\boldsymbol{x}}(\boldsymbol{m}) = G(\boldsymbol{x};\boldsymbol{m},\sigma^2)$} with the appropriate uncertainties for each detection.
This is a two dimensional double integral including all the components in the geometrical model. Lobes depend on the core so the core's position $m$ needs to be established first, followed by the lobes $m'$ and $m^{\prime\prime}$.

Naturally, there are many other competing hypotheses, where we consider only partial matching.
For example, we consider a LOBE-LOBE model in which the putative core is actually an unrelated radio source, we write the LOBE-LOBE likelihood as
\begin{eqnarray}
\label{eqn:linkelikelihood2}
{\cal{}L}_{ll} &=&      \left[\int\!\!d\boldsymbol{m}_0\ p(\boldsymbol{m}_0)\,L_{r_0}(\boldsymbol{m}_0)\right]\!
                 \cdot  \left[\int\!\!d\boldsymbol{m}_1\ p(\boldsymbol{m}_1)\,L_{c}(\boldsymbol{m}_1)\right.\nonumber\\
			    &&      \left. \int\!\!d\boldsymbol{m}_1'\ p(\boldsymbol{m}_1'|\boldsymbol{m}_1)\, L_{r_1}(\boldsymbol{m}_1')\,L_{r_2}(2\boldsymbol{m}_1\!-\!\boldsymbol{m}_1') \right]
\end{eqnarray}
where the first term can be evaluated analytically in case of the flat all-sky prior on $\boldsymbol{m}_0$ and is equal to \mbox{$1\big/4\pi$} because the Gaussian is symmetric in its arguments. The first term is independent of the second, as their host galaxies are not the same, so their integrals are calculated separately, then multiplied~\citep{2008ApJ...679..301B}.

In practice there can be more than three radio sources in close proximity to a given detection in another band, which are considered to be separate sources like the $\boldsymbol{m}_0$. Generalising our model to such complex scenarios will be explored in future work.

A more realistic model would also allow the lobes to be at slightly different distances. We introduce a new parameter $k$ to account for such asymmetry and write
\begin{equation}
\boldsymbol{m}'' = 2\boldsymbol{m} - \boldsymbol{m}' + k(\boldsymbol{m}-\boldsymbol{m}').
\end{equation}
The numerical integration now includes the $k$ parameter over a prior that we choose to be a Gaussian centered on zero, so \mbox{$k\!=\!0$} corresponds to the symmetric case.

%
Each of the radio components can be considered a core, a lobe or a separate object entirely. For example, three radio components $(r_0, r_1, r_2)$ can be CORE-LOBE-LOBE, LOBE-CORE-LOBE, LOBE-NONE-LOBE, NONE-LOBE-LOBE, LOBE-NONE-CORE and so on. We calculate the likelihoods of all these possibilities. The ratio of possible likelihoods for two hypotheses with likelihoods ${{\cal{L}}_1}$ and ${{\cal{L}}_2}$
\begin{equation}
B_{1,2}=\frac{{\cal{L}}_1}{{\cal{L}}_2} \label{eqn:bayes}
\end{equation}
is the Bayes Factor that compares them in an objective way.
When \mbox{$B\!\gg\!1$}, the data clearly prefer the hypothesis in the numerator but when \mbox{$B_{1,2} \approx 0$} the alternative has stronger support. When \mbox{$B_{1,2} \approx 1$} is around unity, more evidence might be needed to make a decision.

\subsection{Implementation Details}
Our catalogs are stored in a database indexed by the Zone Algorithm \citep{2007cs........1171G} to quickly identify possible candidate matches and rule out the completely unlikely ones. We consider lobes with angular separations up to 2 arc minutes. These associations are processed by a custom multi-threaded C\# library that implements the combinatorics and the numerical integration. For a given set of detections we generate all possible hypotheses using the {\em{}Facet.Combinatorics} package by Akison~\footnote{Permutations, Combinations, and Variations using C\# Generics~{http://www.codeproject.com/Articles/26050/Permutations-Combinations-and-Variations-using-C-G}}, and prune the resulting list to only contain the physically possible combinations, e.g., no more than one core. We further exclude models that will clearly perform poorly based on the vicinity of the radio sources to the point source in the other band. This filtering, however, is very conservative and the final decision is based on the probabilistic analysis.

To scale to the large number of possibilities, the numerical integration has to be very fast. Our implementation uses importance sampling  for all integrals \citep[e.g.,][]{numerical1992}. We speed up the calculations by always sampling from the tightest possible Gaussian.
The product of Gaussians and linear parameter transformations yield Gaussians, which makes the procedure an order of
magnitude faster than naive implementations.

The models can contain at most three radio components to fit a core and two lobes. When the number of components is more than 3, e.g., there are 5 nearby radio detections%
,
we consider the extra ones to be from other independent sources. For example, the NONE-LOBE-LOBE-NONE-CORE hypothesis, (where we use the ``NONE" designation to indicate that the respective radio component is not related to the counterpart of interest), will have a likelihood that is the product of an integral identical to the previous ${\cal{L}}_{cll}$ formula with the appropriate data and two independent integrals that are again just \mbox{$1\big/4\pi$}. For the all-NONE hypothesis corresponding to 5+1 independent sources, the likelihood is a constant \mbox{$\left(1\big/4\pi\right)^6$}. We can compare all other hypotheses to this, including those with a different number of components, by considering additional independent sources.

\section{Application to SWIRE and ATLAS}\label{sec:app}
We apply the new algorithm to two well-studied catalogs.
%
The Australia Telescope Large Area Survey (ATLAS; Norris et al. 2006, hereafter Norris06) detected 784 radio components from 726 distinct sources around Chandra Deep Field-South area.
The Spitzer Wide-Area Infrared Extragalactic \citep[SWIRE;][]{2003PASP..115..897L} contains about 113000 objects in the area also covered by ATLAS CDF-S.

\begin{figure}
\begin{center}
\includegraphics[scale=1]{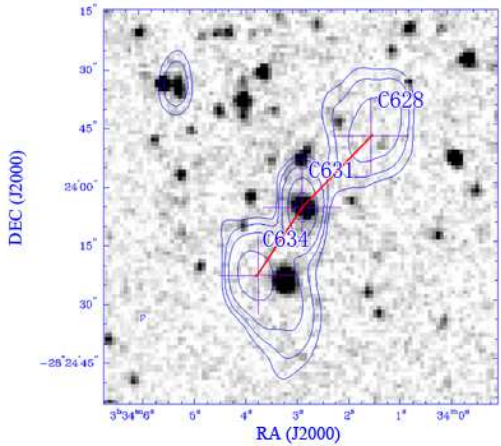}
\end{center}
\caption{A typical cross-identification from Norris et al.(2006). Contours show the radio emission, and greyscale show the SWIRE 3.6~$\mu$m emission. The three radio components (C634,~C631,~C628) were identified by Norris et al.(2006) as a radio triple source hosted by the bright infrared galaxy coincident with C631. \label{fig:c628}}
\end{figure}

We compare our associations to the matched catalog
by \citet{2006AJ....132.2409N}, which is publicly available online.%
\footnote{SWIRE~\&~ATLAS~CDF-S matched catalog {http://cdsarc.u-strasbg.fr/viz-bin/Cat?cat=J/AJ/132/2409}}
These associations were created by expert examination of the radio contours overlayed on the images, as illustrated in Figure~\ref{fig:c628}. There are 119 SWIRE objects that are not available publicly, so we have removed them from further comparison.
The final public sample contains 10 radio triplets and 27 doublets. Some doublets are identified by \citet{2006AJ....132.2409N} as two lobes, while others are identified as a lobe and a core. There are also 558 single sources, which correspond to a single SWIRE source in their list of associations.
Naturally we do not know the underlying physical reality of the published associations but  use these published results as our reference, against which we benchmark our automated method.

\subsection{Most Likely Associations}
To find associations, first we look for the potential radio components around each SWIRE detection using an inclusive search radius of 2 arc minutes.
%
%
This heuristic limit is motivated by previously identified associations in the reference catalog. Its actual value does not affect the quality of the automated associations as long as it finds all candidates, but a large threshold would slow the algorithm down simply because of the large number of candidates to sort through and reject.
Next we evaluate the likelihoods for all possible combinations.
In principle, this is combinatorially expensive and would require a careful analysis of prior probabilities of the different types of matches.
In this study, we apply a simpler algorithm instead to look for the most likely associations.

We use a simple approach to pick the associations based on the calculated likelihoods for the different hypotheses. It is an iterative procedure where we take the association with the largest likelihood. Next we remove from the list of available hypotheses all of those that include the source, which we already accounted for in the best association. The remaining hypotheses are again searched for the best association and we repeat the procedure until nothing is left. This greedy approach runs quickly and provides robust results in our tests.

To compare the different hypotheses one has to characterize the possible geometric arrangements of the cores and lobes. In particular the probability distribution function (pdf) of possible core-lobe distances is needed for the prior. This appears as the conditional density
\mbox{$p(m'|m)$}. This function is expected to rise at small separations but fall at large scales.
One possible such pdf is the Rayleigh distribution. In particular we pick a pdf with a mean of 9 arc seconds, chosen as a reasonable
value that is consistent with the distribution of realistic core-lobe separations in radio images of the sensitivity of ATLAS.
Table~\ref{tbl:recompare} illustrates the quality of our results. At first glance we see that the automated method using only the coordinates and a simple geometric model can find most of the triplets in our references catalog and successfully identify the singletons, too. The doublets are associated with less precision: while the new method misses some of the reference matches, it also identifies extra ones based on the limited information included.
In the next section we will discuss these categories in detail and explore examples to illustrate the scenarios where the reference and automated catalogs differ.

\begin{table}
\begin{center}
\caption{Comparison between the result of automated and manual classification\label{tbl:recompare}}
\begin{tabular}{cccccc}
\hline\hline
Type & Norris06 & Bayesian & Common & Missed & Extra\\
\hline
triplet & 10 & 12 & 9 & 1 & 3 \\
doublet & 27 & 48 & 19 & 8 & 29 \\
single  & 558 & 566 & 536 & 22 & 30 \\
\hline\hline
\end{tabular}
\end{center}
\end{table}

\subsection{Association of Radio Triples}
\begin{figure*}
\begin{center}
\includegraphics[scale=1]{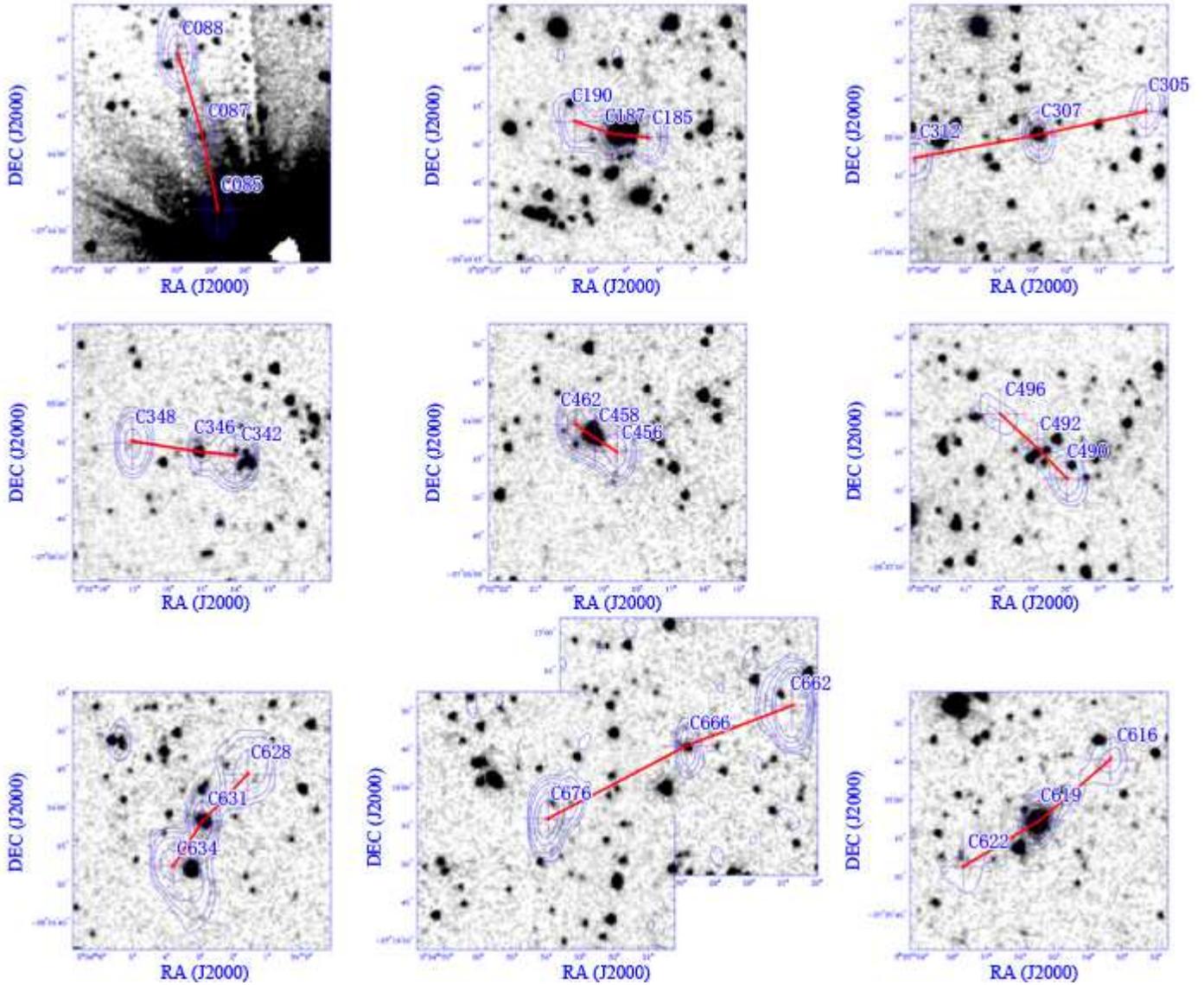}
\end{center}
\caption{An array of sample triplets found both by Norris et al.(2006) and by our method. Each of these triplets has a central core and two lobes nearly in a straight line. \label{3x3}}
\end{figure*}

The automated code finds 9 of 10 of the manually identified triplet associations. Figure~\ref{3x3} shows some of the triplets that were common across the reference and automated catalogs. We see that there are a variety of shapes that are not only often asymmetric in the distances of the lobes from the core but also have a slight angle from the assumed straight model. Our method finds these because the astrometric uncertainty allows sufficient deviation from the straight line. These results have high likelihoods on an absolute scale and also compared to the possible competing scenarios.

\begin{figure}
\begin{center}
\includegraphics[scale=0.6]{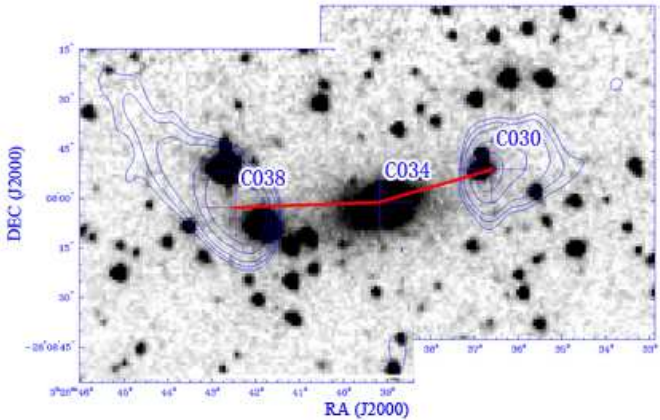}
\end{center}
\caption{A triplet with a large angle between the two lobes. Our program classifies it as a LOBE-LOBE which has larger Bayes Factor comparing to the triplet. \label{c303438project}}
\end{figure}

The only missed triplet has significantly lower likelihood compared with the others. Figure~\ref{c303438project} shows this association. The difference is obvious upon inspection: this geometry is more bent and has an angle of about 15$^{\circ}$ from the straight line. This appears to be too much to be tolerated by the simple model.
A possible improvement of our geometric description would be to include another parameter to capture the bent shape. We plan to explore this possibility in future studies.

\begin{figure*}
\begin{center}
\includegraphics[scale=1]{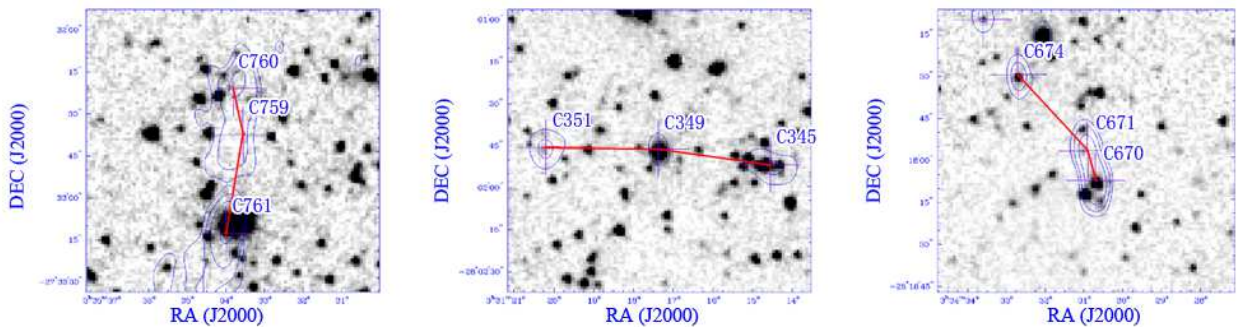}
\end{center}
\caption{Three extra triplets found by the program. Most of them are fit well with the geometry model, but are identified as a line of galaxies or other cases by Norris et al.(2006), see text for detailed discussions. \label{triplet3more}}
\end{figure*}

Our program also finds 3 extra triplets, seen in Figure~\ref{triplet3more}.

Here we discuss these in detail, to guide future refinement of the algorithms.
\subsection*{C761}
The algorithm classifies C760, C759, and C761 as a radio triplet. Norris06 classified them as a  ``linear complex (jets or arc?)''. Both Norris06 and the algorithm associate this source with the SWIRE source $SWIRE3\_J033533.90-273310.9$, and so the only difference is one of terminology, caused by the complex shape of the radio source. Classifying such complex structures is beyond the scope of the present project.
\subsection*{C349}
The algorithm classifies C351, C349, and C345 as a radio triplet. Norris06 classified them as a  ``a line of three galaxies". C351 and C349 are each associated with a catalogued SWIRE galaxy, whilst C351 is associated with an uncatalogued SWIRE source. Given that each of the three radio components is associated with a SWIRE source, Norris06 classified them as there separate galaxies, whereas the algorithm, without access to the uncatalogued data,  found matches for only two of the galaxies, and so classified them as a triplet. If the algorithm had had access to the same catalog as Norris06, then it would have made the same classification.
\subsection*{C371}
The algorithm classifies C670, C671, and C674 as a radio triplet. Norris06 classified them, along with nearby C675, as  ``A line of 4 galaxies". C670 and C674 are each associated with a catalogued SWIRE galaxy, whilst C671 and C675 are associated with uncatalogued SWIRE sources. Again, if the algorithm had had access to the same catalog as Norris06, then it would have made the same classification.

The lesson from this is that the robustness of the cross-matching algorithm, is limited by the availability of multiwavelength data. Norris06 were able to use the SWIRE images which show sources that were absent from the catalog. Modifying the algorithm to enable access to such lower reliability multiwavelength data is
 deferred to future work.

\subsection{Association of Radio Doubles}\label{subsect:associationofradiodoubles}
\begin{figure*}
\begin{center}
\includegraphics[scale=1]{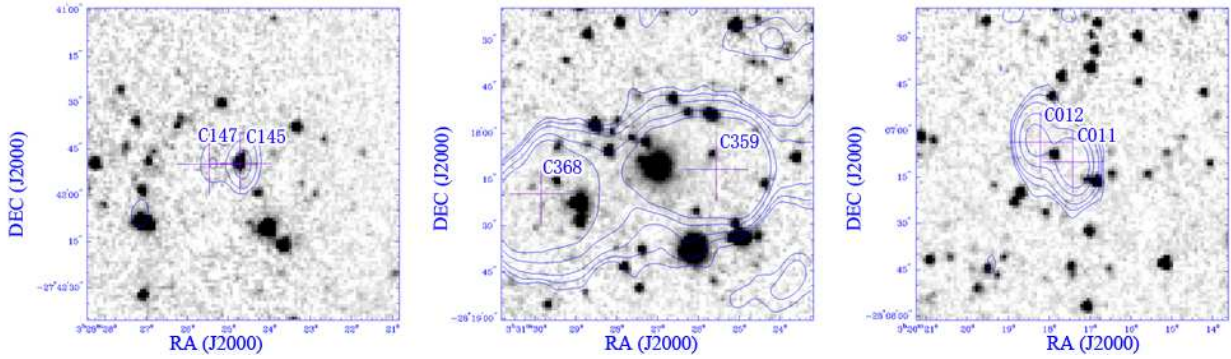}
\caption{Three typical radio doubles found by both methods. The left figure is a CORE-LOBE combination. The middle and the right figures are of LOBE-LOBE type. \label{fig:gooddoubles}}
\end{center}
\end{figure*}

\begin{figure}
\begin{center}
\includegraphics[scale=0.6]{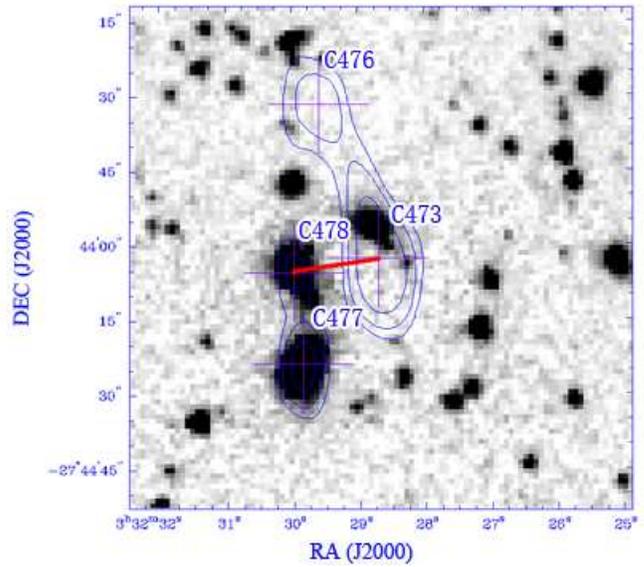}
\caption{A radio double that our program did not find. The algorithm identified the radio objects C478 \& C473 as radio double, and this association has a larger Bayes Factor than the correct pairs, as discussed in Section~\ref{subsect:associationofradiodoubles}.
\label{fig:baddoubles}}
\end{center}
\end{figure}

Among the 19 doubles that are common in the reference and automated catalogs, 14 are LOBE-LOBE combinations. Three typical examples are shown in Figure~\ref{fig:gooddoubles}.
There are eight doubles that our code does not find.
In four of these cases the automated code associates the radio pairs with other SWIRE objects, in combinations that appear to have larger likelihoods.
In the four remaining cases, the code splits the radio components, and rearranges them with other radio detections.
Figure~\ref{fig:baddoubles} shows one of these misidentified associations, which we now discuss in detail as a case study.

The algorithm associates C473 and C478 as a double, and classifies C477 and C476 as isolated galaxies. Norris06 identify C477 and C478 as a pair of galaxies, and C473 and C476 as a double-lobed radio source, because there is a low-brightness bridge of radio emission between C473 and C476, so these are likely to be associated, and between C477 and C478, so these are likely to be associated. Furthermore, C477 and C478 each have a bright IR counterpart, so are likely to be galaxies. Neither C473 nor C476 have an IR source coincident with the radio peak, but there is a bright IR galaxy part-way along the bridge, which is likely to be the host, especially since the bridge is bright at that position.

To improve the algorithm, greater weight needs to be given to the presence of  IR counterparts (such a C477 and C478). More importantly, visual classification is advantaged by the potential to
make use of extended or low signal-to-noise features, such as the faint extensions from the radio source peaks that link
the lobes in these examples. This information is not captured by the simple positional information retained in a source catalog, limiting
the performance of the automated algorithm in this situation. It is difficult to see how such information can readily be incorporated into a Bayesian algorithm.

Among the 29 extra doubles found by the automated procedure, there are 14 LOBE-LOBE combination and the rest are all CORE-LOBE.
While many of these may be real, some are clearly two radio components sitting within an extended SWIRE object, resulting their LOBE-LOBE or CORE-LOBE hypotheses having higher likelihoods than being isolated sources.
Although this leads to the classification of such sources as radio doubles, they are unlikely to be so in reality. Ensuring the automated process can recognize and exclude this kind of false-double is part of the planned refinements.

\section{Conclusion}\label{sec:summary}
We have introduced a novel approach to automatically crossmatch radio sources having a realistic morphology against point source counterpart catalogs.
Instead of the usual catalog matching, where detections in  separate datasets are simply assigned to each other when co-located, we can also identify components of the radio sources as cores or lobes separated from each other on the sky. Traditionally such matching could only be done manually by astronomers through visual inspection of the overlays of images.

The new method was tested on the SWIRE infrared and ATLAS radio CDF-S catalogs.
{We find that the automatic associations agree well with the reference catalog. The differences all arise either from inadequate representation of the data by the catalog, or from the simplicity of our current radio source morphology model, which only allows for the components to lie along a straight line.}
Also the current procedure only uses the directions of the measured sources, while experts constructing a manual catalog would typically rely on shape and brightness information as well. This study is a first step toward automating a manually intensive process in order to provide a scalable functionality for the coming generation of massive radio surveys with Square Kilometre Array pathfinder telescopes.

\section*{Acknowledgments}\label{ack}
{This project was germinated in the inspiring atmosphere of the 2011 Lorentz Workshop on ``Probing the Radio Continuum Universe with SKA Pathfinders" for which the authors are extremely grateful to the organizers and the Lorentz Center.
TB and DF acknowledge support from the Virtual Astronomical Observatory under the National Science Foundation cooperative agreement AST-0834235.
DF was partially supported by National Natural Science Foundation of China(U1231108) and Chinese Academy of Sciences (XXH12503-05-05).}

\label{lastpage}
\end{document}